# DYNAMIC THERMAL ANALYSIS OF A POWER AMPLIFIER

*Jedrzej Banaszczyk* *[#], *Gilbert De Mey* [#], *Marcin Janicki* *, *Andrzej Napieralski* *,
*Bjorn Vermeersch* [#], *Piotr Kawka* [# †]

\* Department of Microelectronics and Computer Science, Technical University of Lodz,
Al. Politechniki 11, 90-924 Lodz, Poland
[#] Department of Electronics and Information Systems, University of Ghent,
Sint Pietersnieuwstraat 41, 9000 Ghent, Belgium
[†] Agilent Technologies, Pegasus Park, De Kleetlaan 12A bus 12, 1831 Diegem, Belgium

## ABSTRACT

This paper presents dynamic thermal analyses of a power amplifier. All the investigations are based on the transient junction temperature measurements performed during the circuit cooling process. The presented results include the cooling curves, the structure functions, the thermal time constant distribution and the Nyquist plot of the thermal impedance. The experiments carried out demonstrated the influence of the contact resistance and the position of the entire cooling assembly on the obtained results.

## 1. INTRODUCTION

The increased density of power dissipated in the state-of-the-art electronic devices and circuits has forced engineers to seek for new and innovative ways of thermal analysis of electronic structures. Since power circuits often operate in a pulsed mode, their steady state thermal conditions are practically never reached. Therefore, in the case of power amplifier analyzed here it is important to determine not only the steady state temperature distribution, but also its dynamic thermal behaviour. Then, knowing the transient characteristics it is possible to obtain a useful dynamic thermal model.

The next section provides a brief overview of various methods of dynamic thermal analysis of electronic circuits such as: cooling curve, thermal time constant distribution, differential and cumulative structure function and finally the Nyquist plot of thermal impedance. Next, the theory is used in the experimental part of the paper to investigate, based on the example of a power amplifier, the effects of varying the contact resistance between the amplifier package and the cooling fin. This was done by loosing the screw or applying thermal grease. Moreover, the influence of the cooling fin position has been also investigated.

## 2. DYNAMIC THERMAL ANALYSIS

There exist several methods of representing the dynamic thermal behaviour of electronic structures. One of such methods is the structure function approach. This approach had been originally developed by Szekely and his research team. For thermal analyses in electronics, the differential structure functions are of a particular interest because they provide the incremental thermal capacitance as a function of the distance from the junction (heat source). The main advantage of this representation is that, as it will be shown later on, it allows the direct identification of particular sections of the heat path of an electronic structure, such as chip, package, solder or radiator. More theory on this subject can be found in [1]-[3].

Another possible method of describing the dynamic behaviour is to use Nyquist plots of the complex thermal impedance. The thermal impedance as a function of the angular frequency $\omega$ can be found performing the Laplace transform of the structure transient thermal response [4]. Owing to this, in most cases it is possible to determine the time constant distribution analytically because the Nyquist plots resemble a combination of circular arcs. As it was proved in [5], for simple geometrical shapes the Nyquist plot can be fitted to a circular arc with a high precision.

For all the above described thermal analysis methods, it is required to take adequate temperature measurements. The measurements can be taken using a forward biased p-n junction. If the bias current is constant, the voltage drop across the junction is the measure of temperature. Usually, temperature is measured as the response to the power step excitation. However, this requires using the same junction for heating and measurement, which might cause some technical problems. An alternative method, used also in this paper, is to heat the junction till steady state is reached and to switch the power off. In this way a cooling curve is obtained, which is just the complement of the heating curve and contains the same information.

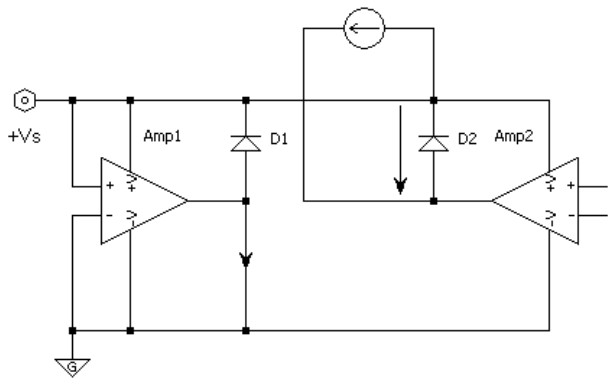

*Fig. 1: Simplified schematics of the amplifier.*

## 3. MEASUREMENTS

This section presents in detail the thermal analyses of the power amplifier. First the circuit and the entire measurement setup is introduced. Then, the measurement results are presented and discussed.

### 3.1. Measurement setup

For the experiments the APEX PA60 power amplifier was used. The circuit, whose simplified schematics is shown in Fig. 1, was chosen expressly to simplify the measurement procedure. This was possible owing to the fact that the package contains twin operational amplifiers with their outputs protected by diodes connected between the output and the positive power supply as shown in Fig. 2. Then, one of the amplifiers was powered in order to provide the heat dissipation and the protection diode and the other one was used as the temperature sensor. Consequently the temperature measurement point does not coincide exactly with the heat source. Hence, there was an inevitable delay between the sensed temperature and the temperature of the heated operational amplifier. However, because both amplifiers are integrated on one single chip, the delay might be assumed to be very small. Experimentally it was found that the delay time did not exceed 100 μs. Finally, as shown in Fig. 2, the power amplifier has been screwed to a cooling fin, which in turn was placed on a thermally insulating material. All the dimensions in the figure are given in mm.

The measurements were performed using the thermal tester T3Ster provided by the MicRed company. The heating amplifier was operated in the saturation mode with the constant power dissipation of 10 W till steady state temperature was reached. After switching the power off, the protection diode of the other amplifier was used to measure the temperature cooling curve. The forward current forced through the diode during the measurement was 1mA. The photograph of the entire equipment used for the measurements is shown in Fig. 3.

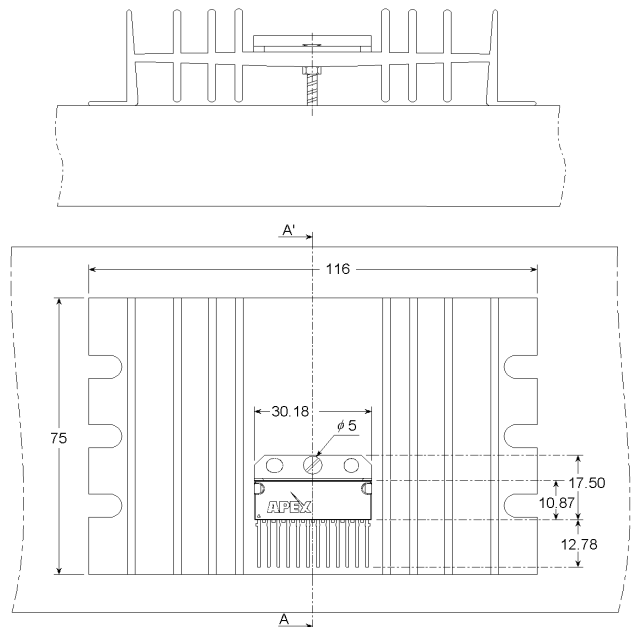

*Fig. 2: View of the amplifier with the radiator.*

### 3.2. Interface thermal resistance influence

Initially, the measurements were taken for the radiator placed horizontally in three different configurations. First the radiator was firmly attached to the circuit. Then, the screw fixing the package to the cooling fin was loosed slightly, thus introducing some additional thermal contact resistance. Finally, thermal grease was applied and the screw was tightened again to improve the thermal contact. The captured cooling curves are presented in Figure 4. The logarithmic time scale was used so as to visualize better the very beginning of the cooling process, because after some 90 s the cooling curves are indistinguishable. As can be seen in the figure, tightening the screw and the use of thermal grease lowered the total thermal resistance by more than 1 K/W (the power dissipated was 10 W).

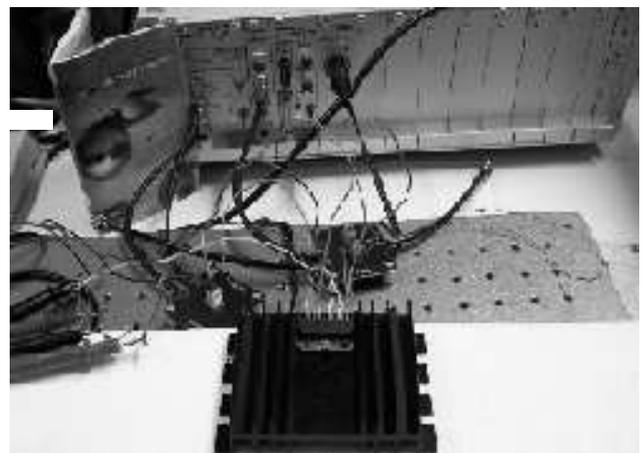

*Fig. 3: Photograph of the equipment.*

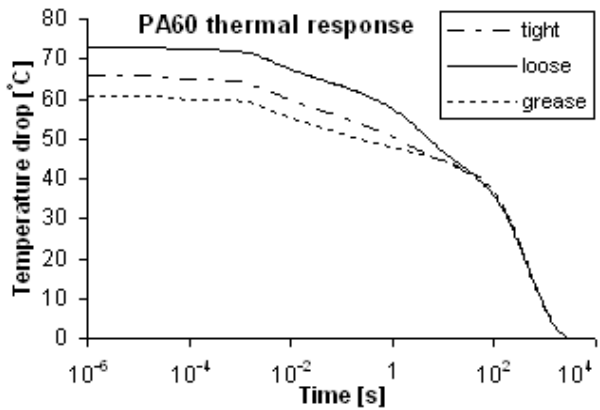

Fig. 4: Cooling curves - influence of interface resistance.

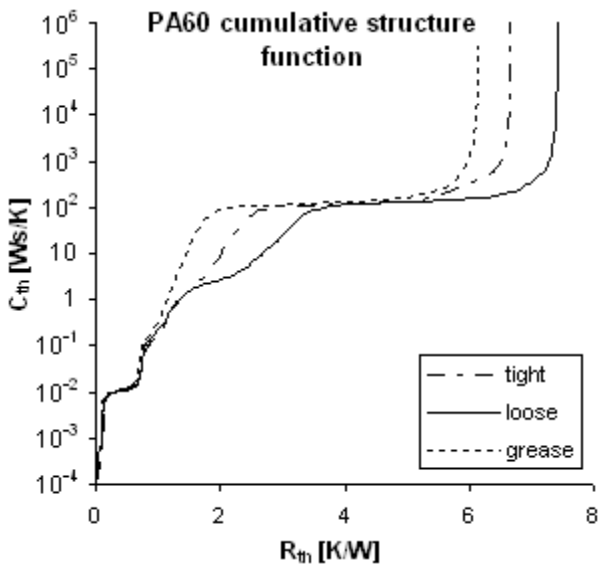

Fig. 5: Cumulative structure functions
- influence of interface resistance.

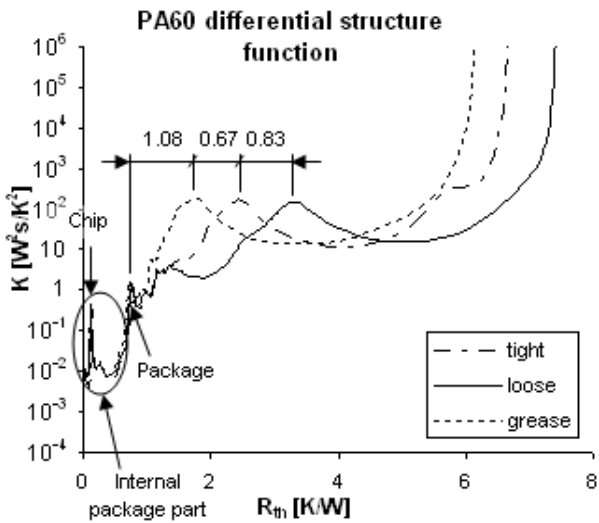

Fig. 6: Differential structure functions
- influence of interface resistance

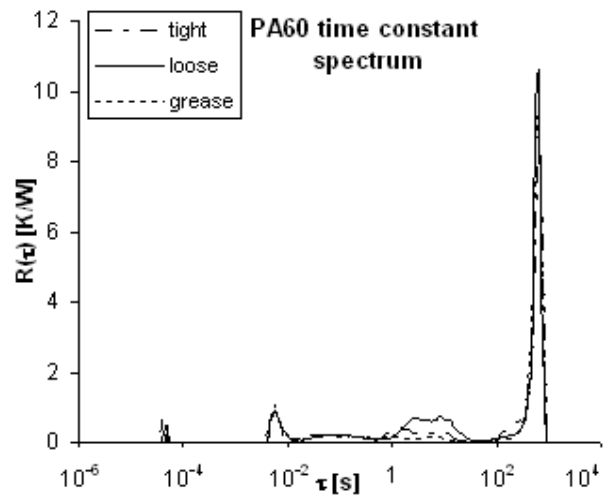

Fig. 7: Time constant distribution
- influence of interface resistance.

The measured cooling curves were processed further using the software provided by MicRed company together with the measurement equipment. The results of various thermal analyses performed by the authors are presented in Figures 5-8, which show respectively: the cumulative structure function, the differential structure function, the thermal time constant distribution and the Nyquist plot of thermal impedance. The cumulative structure functions are useful in the steady state thermal analysis, whereas the other quantities are important in the dynamic analysis.

Examining the cumulative structure function pictured in Figure 5, it can be observed that the change of contact resistance does not affect the total value of the thermal capacitance (the large plateau is located at the same value of the thermal capacitance) but it influences the thermal resistance (see the shift of the vertical lines at the ends of the curves). Namely, tightening the screw lowered the total thermal resistance by 0.8 K/W and the application of the thermal grease reduced its value further by 0.6 K/W down to 6.2 K/W.

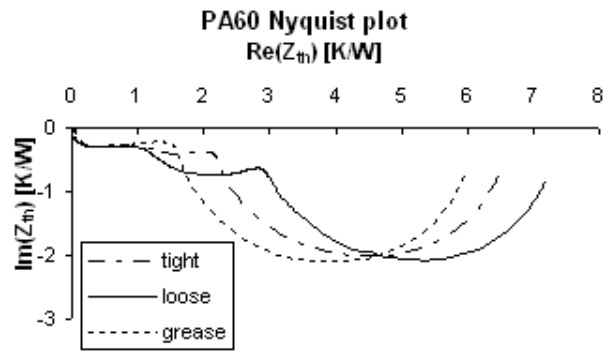

Fig. 8: Thermal impedance
- influence of interface resistance.

Investigating the differential structure function, shown in Figure 6, further interesting observations can be made. In this figure there are three peaks visible. These peaks can be attributed to the chip, the package, and the cooling fin. The shift of the third peak caused by the change of the interface resistance is evident. The distance between the second and the third peak can serve as the measure of the thermal resistance of the contact between the package and the radiator. As indicated in the figure, the value of this resistance for tightly screwed radiator with thermal grease slightly exceeds 1 K/W, whereas in the case of loosely attached cooling fin it amounts to almost 2.6 K/W.

Furthermore, the time constant distribution presented in Figure 7 clearly shows that there are four peaks located near the values of 60µs, 9ms, 8s and 800s. These peaks correspond to the four aforementioned sections of the heat flow path, i.e. the chip, the package, the interface and the radiator. The variation of the interface thermal resistance affects only the peak corresponding to the time constant of a few seconds. This peak has almost disappeared when the thermal grease was used, so it has to be related to the contact thermal resistance.

The Nyquist plot of the thermal impedance, presented in Figure 8, shows three circular arcs centered around the same time constant values as the last three peaks in the time constant distribution. The high frequency peak is not visible on the Nyquist plot. For the well-attached radiator, the middle arc in the Nyquist plot practically disappears similarly to the peak in the time constant distribution, thus proving that this circle can be attributed to the interface thermal resistance.

Moreover, it is worth noticing that the change of the interface resistance does not affect the overall shape of the analyzed curves in all the regions reflecting the heat flow through the inner parts of the package and the radiator.

### 3.2. Cooling condition influence

The second part of the experiments presented here was devoted to the investigation of the effects of the radiator position on the previously analyzed curves. The position of the radiator was changed by varying its tilt angle. Except for the horizontal position, the tilt angles of 45° and 90° (vertical position) were considered. Finally, in the last experiment the radiator was lifted by some 5 mm, thus creating an air gap between the radiator and the insulating material, but still maintaining the horizontal position. All the measurements were performed for the tightly screwed radiator without the thermal grease. The obtained results of thermal analyses are shown in Figures 9-13. Similarly to the previous cases, the charts represent: the cooling curves, the cumulative structure function, the differential structure function, the time constant distribution and the Nyquist plot of the thermal impedance.

The examination of the curves shown in Figures 9-11 provides some important conclusions. Namely, when the radiator was slightly tilted, the hot air from beneath could be evacuated more efficiently by the gravitational forces through the natural convection thus reducing the thermal resistance. The change of the position from the horizontal one to 45° caused the reduction of the thermal resistance by 0.6 K/W, which constitutes almost 10% of the total junction-to-ambient resistance. Further tilting the radiator to the vertical position did not influence the value of the thermal resistance significantly. On the other hand, lifting of the radiator produces the decrease of 0.4 K/W owing to the improved cooling from the bottom.

Analysing Figures 12-13, it can be observed that the cooling conditions have no impact on the high frequency time constants. Thus, when the time constant distribution is concerned, only the large peak around 800 s is changed. Hence this peak can be attributed to the heat transfer from the radiator to the ambient by convection. It is also worth noticing that the third peak corresponding to the thermal interface between the package and the cooling fin was not influenced by these experiments at all. Similarly, in the Nyquist plot, only the largest low frequency circular arc was influenced. Concluding, compared to the previous experiment, the cooling conditions changed the analyzed curves only in the regions representing the heat transfer between the radiator and the ambient without affecting their shape in all the regions reflecting the heat flow from the heating junction to the radiator.

### 4. CONCLUSIONS

In this paper, the thermal dynamic behaviour of a power amplifier was discussed thoroughly employing different analysis methods. Specifically, the effects of the interface thermal resistance and the radiator position on the circuit temperature were investigated. The influence of both phenomena was illustrated in various diagrams. Analysing the results it was possible to characterize thermally the entire structure. Each part of the resulting curves was attributed to different sections of the heat path between the junction and the ambient. The presented methodology can be applied for various practical investigations such as compact model creation or fault identification, e.g. for die attach delamination detection.

### ACKNOWLEDGEMENTS

Bjorn Vermeersch, a Research Assistant for the Scientific Research Foundation - Flanders (Fonds Wetenschappelijk Onderzoek - Vlaanderen), would like to thank for their financial support. This research was supported also by the internal grant of the Technical University of Lodz Dz.St. K-25/1/2006.

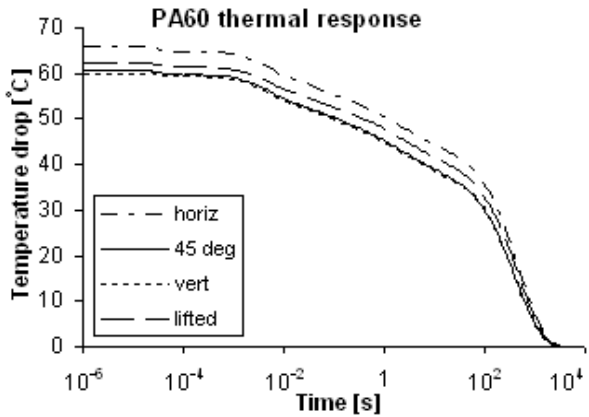

*Fig. 9: Cooling curves - position influence.*

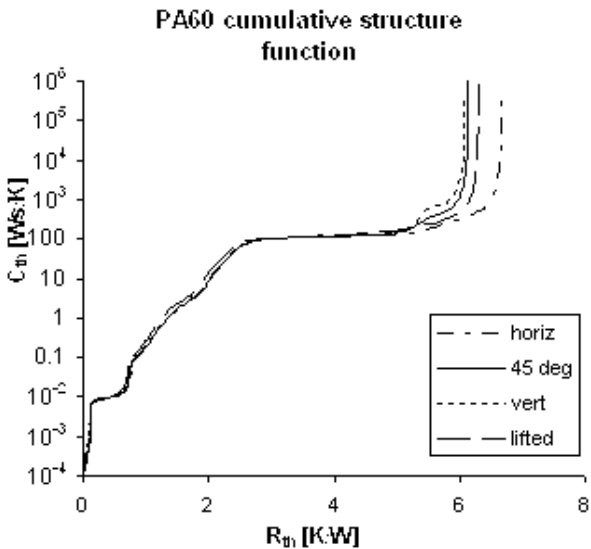

*Fig. 10: Cumulative structure function - position influence.*

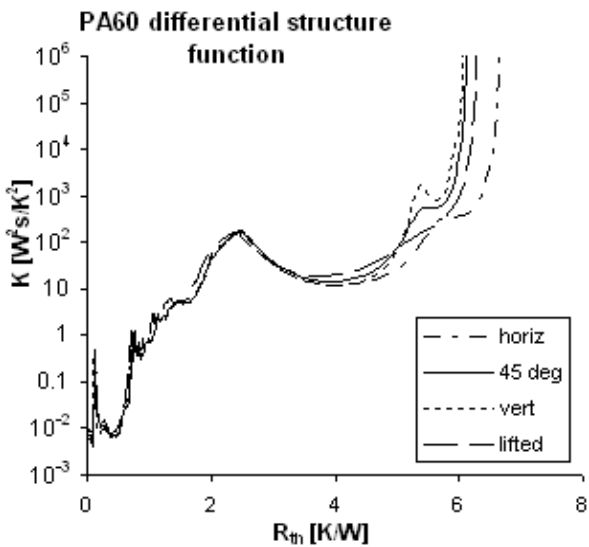

*Fig. 11: Differential structure function - position influence.*

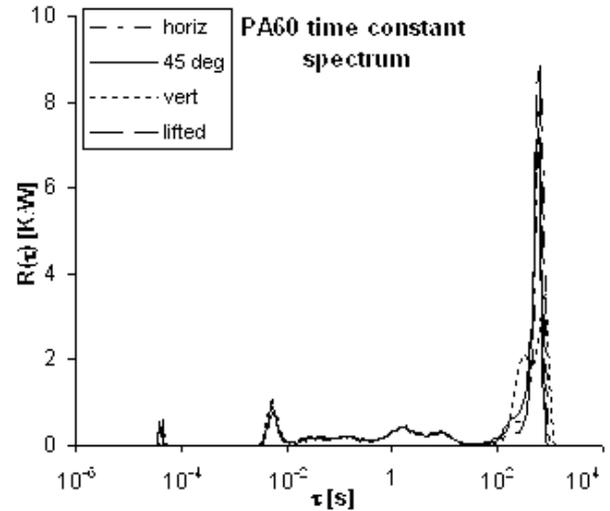

*Fig. 12: Time constant distribution - position influence.*

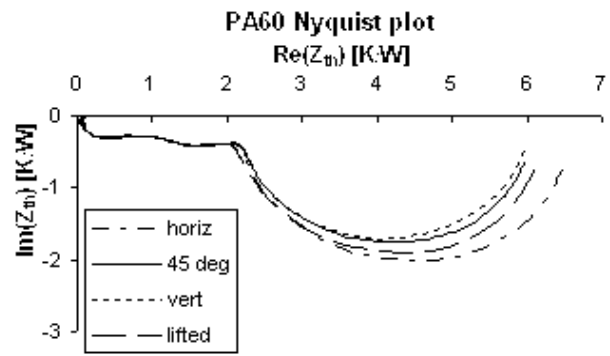

*Fig. 13: Thermal impedance – position influence.*